\documentstyle[twocolumn,prl,aps,psfig]{revtex}
\tightenlines
\begin{document}
\draft
\title{Electronic spin precession in semiconductor quantum dots
with spin-orbit coupling}
\author{Manuel Val\'{\i}n-Rodr\'{\i}guez, Antonio Puente,
and Lloren\c{c} Serra}
\address{Departament de F\'{\i}sica, Universitat de les Illes Balears,
E-07071 Palma de Mallorca, Spain}
\author{Enrico Lipparini}
\address{Dipartimento di Fisica, Universit\`a di Trento,
and INFM sezione di Trento, I-38050 Povo, Italy}
\date{May 9, 2002}
\maketitle
\begin{abstract}
The electronic spin precession in semiconductor dots is strongly affected 
by the spin-orbit coupling. We present a theory of the electronic
spin resonance at low magnetic fields
that predicts a strong dependence on the dot 
occupation, the magnetic field and the spin-orbit coupling strength. 
Coulomb interaction effects are also taken into account
in a numerical approach.
\end{abstract}


In recent years the spin properties of non-magnetic semiconductors
have attracted an increasing attention, not only for the fundamental 
physics behind the subject but also for the future technological 
applications of the electronic spin in spin-based devices \cite{Wol01}.
The available experimental techniques allow for a precise observation 
of spin dynamics in a wide range of semiconductor structures. 
Actually, spin precession can be monitored with femtosecond resolution 
using time resolved Faraday rotation, as reported
in Ref.\ \cite{Sali01} for GaAs quantum wells; and in 
Ref.\ \cite{Gup99} for CdSe excitonic quantum dots. 
Another exciting possibility comes from spatially resolved spin
detection, achieved in Ref.\ \cite{Dur02} for organic molecules
on a graphitic surface by combining the spatial resolution of 
scanning-tunneling microscopy (STM) with the spin sensitivity
of electron-spin resonance (ESR).  

In this work we report a theoretical study of the spin precessional 
properties of electrons confined to a model GaAs quantum dot, including
spin-orbit (SO) coupling. This mechanism gives rise to a
rich variety of spin precessional frequencies, depending on the 
orbital state of the electrons, even in the absence of a vertical 
magnetic field; as compared to the Larmor frequency for systems 
without SO coupling. 

In order to model a GaAs quantum dot with SO coupling for the standard
(001) plane \cite{Aro93}, we add to the Hamiltonian 
the Dresselhaus term originating from the bulk inversion asymmetry:
\begin{equation}
\label{eq1}
{\cal H}_D = \frac{\lambda_D}{\hbar} \sum_{i=1}^{N}{ 
\left[\, P_x\sigma_x-P_y\sigma_y\,\right]_i} \; ,
\end{equation}  
where the $\sigma$'s are the Pauli matrices and 
${\bf P}=-i\hbar\nabla+\frac{e}{c}{\bf A}$ represents the canonical 
momentum containing the vector potential ${\bf A}$ 
---within the symmetric gauge for a vertical magnetic field $B$
one has ${\bf A}=\frac{B}{2}(-y,x)$.
The intensity of the SO term depends on the effective dot height $z_0$
as \cite{Vos01}
$\lambda_D\approx\gamma(\pi/z_0)^2$
where $\gamma$ is a material dependent constant that for GaAs 
takes the value $\gamma=27.5$~eV{\AA$^3$} \cite{Kna96}.
In the present work we shall consider 
$\lambda_D$ parameters in the
range $[\,0.44\,,\,1.08\,]\times 10^{-9}\;{\rm eV}{\rm cm}$, in the 
same order of magnitude of those found in the literature 
for GaAs/AlGaAs heterojunctions \cite{Vag98} 
($0.25\times10^{-9}\;{\rm eV}{\rm cm}$ for electrons,
 $0.6\times10^{-9}\;{\rm eV}{\rm cm}$ for holes).

In the following, the dot vertical extent only determines the 
SO coupling strength, the electronic motion is otherwise considered
bidimensional in a lateral confinement potential with circular 
symmetry, $V(r)$. Taking into 
account the Zeeman ${\cal H}_Z$ and spatial  
${\cal H}_0$ energies and neglecting for the moment electron-electron
interaction, the full Hamiltonian reads 
${\cal H}={\cal H}_0+{\cal H}_D+{\cal H}_Z$, where
\begin{eqnarray}
\label{eqH0}
{\cal H}_0 &=& \sum_{i=1}^{N}{
\left[\rule{0cm}{0.5cm}\right.
\frac{{\bf P}^2}{2m^*}+V(r)
\left.\rule{0cm}{0.5cm}\right]_i}\; ,  \\
\label{eqHZ}
{\cal H}_Z &=& \sum_{i=1}^{N}{
\left[\rule{0cm}{0.5cm}\right.
\frac{1}{2} g^* \mu_B B \sigma_z
\left.\rule{0cm}{0.5cm}\right]_i}\; .
\end{eqnarray}
In Eqs.\ (\ref{eqH0}) and (\ref{eqHZ}) 
$\mu_B$ denotes the Bohr magneton while 
$m^*$ and $g^*$ are, respectively, the 
effective mass and gyromagnetic factor for the conduction 
band of bulk GaAs, i.e., $m^*=0.067\, m_e$ and $g^*=-0.44$.

For ${\cal H}_0 \gg {\cal H}_D \gg {\cal H}_Z $
a diagonalization in spin space 
to order $\lambda_D^2$ can be obtained by means of a unitary
transformation \cite{Ale01} to a new Hamiltonian
$\tilde{\cal H}= U^+{\cal H}U$, with
\begin{eqnarray}
\label{eq3}
\tilde{\cal H} &=& \sum_{i=1}^N{ 
\left[\rule{0cm}{0.5cm}\right.
{{\bf P}^2\over 2m^*} 
+ V(r)}
+ \lambda_D^2 \frac{m^*}{\hbar^3}\, (xP_y - yP_x) \sigma_z \nonumber\\
&+& \frac{1}{2}\, g^* \mu_B B \sigma_z
\left.\rule{0cm}{0.5cm}\right]_i 
-N \lambda_D^2\frac{m^*}{\hbar^2}
+ O(\lambda_D^3)\; .
\end{eqnarray}
In the new, {\em intrinsic}, reference frame the eigenstates are 
orbitals with well defined spin and spatial angular
momentum in $z$ direction, i.e., 
$\varphi_{n\ell\pm}({\bf r})\chi_\pm(\eta)$; where 
$\eta=\uparrow,\downarrow$ and 
$\sigma_z\chi_\pm=\pm\chi_\pm$.
Note that the spatial parts also depend on the spin label since the 
effective radial confinement in Eq.\ (\ref{eq3}) is different for 
$\chi_+$ and $\chi_-$ orbitals.
When these eigenstates are transformed back to the laboratory 
frame, spin and angular momentum become ill defined, but they 
deviate little from the well defined intrinsic values. Therefore,
we shall retain the intrinsic labels $(n\ell\pm)$
to characterize the laboratory-frame eigenstates
\begin{eqnarray}
\label{eq5}
\chi_{n\ell+}({\bf r},\eta) &\equiv&
\varphi_{n\ell+}({\bf r})
{1 \choose -i\lambda_D\frac{m^*}{\hbar^2} re^{-i\phi}}\nonumber\\
\chi_{n\ell-}({\bf r},\eta) &\equiv&
\varphi_{n\ell-}({\bf r})
{-i\lambda_D\frac{m^*}{\hbar^2} re^{i\phi} \choose 1}\; .
\end{eqnarray}
Using this approach, in Ref.\ \cite{Val02a} we showed that the static 
spin of 
odd-$N$ quantum dots alternates between up and down states as a consequence 
of the SO coupling when the magnetic field and/or the SO coupling strength 
are varied. Other static properties have been studied by Governale
\cite{Gov02} using a SO coupling of the Rashba type,
unitarily equivalent to the Dresselhaus contribution \cite{Ale01}.  
Here we shall focus on the dynamical spin 
evolution in a model quantum dot when all the system parameters are kept 
fixed.

In order to excite the electronic spin precession one needs to perturb 
the ground state spin configuration. The usual way to achieve this 
consists in  
applying a horizontal magnetic field for a certain time interval
that rotates the spin and triggers the precessional motion.
By performing a spin rotation about an arbitrary horizontal axis 
of the above given spinors and decomposing the result
in the stationary basis (\ref{eq5}), we observe some interesting features.
In absence of SO coupling ($\lambda_D=0$) the only allowed
transitions are the spin flips $(n\ell+)\leftrightarrow(n\ell-)$,
leading to an in-plane spin precession at the usual Larmor frequency
$\omega_L=|g^*|\mu_B B/\hbar$. 
This is a well-known result valid even when spin-independent  
interactions are present \cite{Sli90}. 
When SO coupling is considered, besides the pure spin flips other 
transitions involving additional changes in $\ell$ and/or $n$
are allowed. In addition to monopolar $\delta\ell=0$, dipolar 
$\delta\ell=\pm1$ and 
quadrupolar $\delta\ell=2$ spin flip transitions also contribute 
with different weights. As we shall show below the $\delta\ell=0$ 
transitions of the pure Larmor mode are still the dominant ones in 
the precessional spectrum with SO coupling; the dipolar ones are weaker 
by more than an order of magnitude while the quadrupolar spin flip excitations 
turn out to be negligible in all cases studied. It is worth to point
out that even transitions between orbitals with different $n$ could 
contribute because of the non orthogonality of $+$ and $-$ 
radial functions, although these will normally involve high energies
and low strengths due to the small deviation from pure
orthogonality. 

To quantify the SO coupling effect we shall assume a parabolic confinement
potential whose eigenenergies and eigenstates are analytically 
known:\cite{Foc28}.
\begin{equation}
V({\bf r})= \frac{1}{2} m^* \omega_0^2 (x^2 + y^2)\; .
\end{equation}
With the above analysis the lowest $\delta\ell=0$ spin-flip mode,
that we shall call the SO precessional mode $\omega_{\cal P}$, has
a frequency, in the low $\lambda_D$ limit, 
\begin{eqnarray}
\label{eq7}
\omega_{\cal P} =
\left|\rule{0cm}{0.75cm}\right.
\omega_L &+& 2\, \ell\, \lambda_D^2 \frac{m^*}{\hbar^4}\nonumber\\
&-& (2n+|\ell|+1) 
{\lambda_D^2 \omega_c \over 
\sqrt{\omega_0^2+\frac{\omega_c^2}{4}}}\frac{m^*}{\hbar^4}
\left.\rule{0cm}{0.75cm}\right| \; .
\end{eqnarray}
The SO precessional mode $\omega_{\cal P}$ is the dominant excitation 
in the spin rotation spectrum and it can be considered in a natural way
as the modification of the pure Larmor mode $\omega_L$
by the SO coupling. Note that in Eq.\ (\ref{eq7}) we have introduced the 
usual cyclotron frequency $\omega_c=eB/(m^*c)$ and that $\ell$ is the 
angular momentum in the intrinsic frame
of the precessionally active orbital, i.e., that with spin flips allowed
by the Pauli principle; normally the 
highest or the second highest occupied level in an odd-$N$ dot.
Even $N$ systems will generally not possess a net spin at low
enough magnetic fields ${\cal H}_D\gg{\cal H}_Z$. 

Equation (\ref{eq7}) already allows us to point out several interesting 
predictions: 
a) At $B=0$ the SO precessional frequency does not
vanish when $\ell\ne0$, with the offset indicating the SO coupling 
intensity.  
b) In general, positive and negative $\ell$ orbitals will display
different $B$ dispersions for a fixed $\lambda_D$.
c) When the precessionally active orbital
changes due to an internal rearrangement the SO precessional 
frequency will display a discontinuity.

It is worth to mention that a $B=0$ offset similar to the one mentioned 
above was observed in GaAs quantum wells already in 1983 \cite{Ste83} 
and that a zero-field level splitting in quantum dots was also discussed 
in Ref.\ \cite{Vos01}. Note that since the SO coupling term is time 
reversal invariant Kramers degeneracy is preserved at $B=0$ and, therefore,
the precessional offset can be found only when the transition involves 
non-conjugate states.

The validity of the preceding $O(\lambda_D^2)$ analysis 
can be tested with direct numerical calculations which 
avoid the approximate diagonalization procedure.
To this end, we have implemented in a spatial grid the solution
to the time-dependent Schr\"odinger equation, labelling the discrete set 
of orbitals by an index $j$,  
\begin{equation}
\label{eq8}
i\hbar\frac{\partial}{\partial t} \chi_j({\bf r},\eta) =
\sum_{\eta'}{
h_{sp}({\bf r},\eta\eta')\; \chi_j({\bf r},\eta')} \; ;
\end{equation} 
where we have defined the single-particle Hamiltonian from
\begin{equation}
\label{eq9}
{\cal H}\equiv\sum_{i=1}^{N}{h_{sp}({\bf r}_i,\eta_i\eta'_i})\; .
\end{equation}
A `real time' simulation of the precession can be performed by taking 
the stationary solutions to Eq.\ (\ref{eq8}), rotate in spin space
with a given horizontal axis (say the $x$ axis)
and use the resulting perturbed spinors as starting point for the 
time evolution. Figure 1 displays one such simulation for a 
$N=7$ quantum dot. The analysis is based on the $x$-component of the total 
spin, in time (lower panel) and energy (upper panel) domains.
The different features discussed above in the analytical model are 
nicely manifested by the numerical signals. From Fig.\ 1 we can  
also see quantitatively the strength of the SO precessional mode with 
respect to the doubly split upper and lower branches of the dipole modes.
The minor energy differences between the numerical and analytical 
peak positions can be attributed to effects beyond $O(\lambda_D^2)$
and, also, to a slight departure
from the ${\cal H}_Z\ll{\cal H}_D$ limit for a finite $B$.

A shortcoming of the time simulation technique is found in the 
determination of very low energy excitations. Low frequency signals
may require extremely long simulation times, exceeding the 
limit of computational feasibility; either by excessive computing time
or by accumulated numerical error. In our case, we have estimated this 
limit at $T_{\em max}\approx 100\, {\rm ps}$ and the 
corresponding minimum frequency 
$\omega_{\em min}/2\pi \approx 10\, {\rm GHz}$. 
Nevertheless, in the noninteracting
case one can directly compute the {\em perturbative strength} function
from the stationary ground state,
\begin{eqnarray}
\label{eq10}
S_{\em prec}(\omega) &=& \sum_{ij}{
(1-f_i)f_j\, 
\left|\langle \chi_i | \sigma_x | \chi_j \rangle\right|^2} \nonumber \\
&\times& \delta(\varepsilon_i-\varepsilon_j-\hbar\omega) 
\; ,
\end{eqnarray}
where $i$ and $j$ span the whole single particle set while the $f_i$'s 
and $\varepsilon_i$'s give
the orbital occupations and energies, respectively. 
We have checked that the perturbative and time simulation methods yield
the same results when both are feasible, whereas the sub-{10 GHz} 
points in Fig.\ 2 have been computed using Eq.\ (\ref{eq10}).

A systematics of the lowest peak energy, i.e., the SO precessional mode,
is gathered in Fig.\ 2 as a function of the magnetic field 
for two different $\lambda_D$'s. Focussing 
first on the non-interacting results, we note that
for the smaller $\lambda_D$ the 
agreement between analytical and numerical values is excellent, 
proving the equivalence of both methods; while the slight differences
for $\lambda_D= 1.1\times 10^{-9} {\rm eV} {\rm cm}$ can be 
understood on the basis of the previous discussion.
The already mentioned offset at $B=0$ with respect to the Larmor 
frequency is clearly seen in 
Fig.\ 2 for $N=7$ and 11, as well as the diferent slopes for different $\ell$
and $\lambda_D$ values.

A better understanding of the precessional mode systematics is obtained
from Fig.\ 3, which displays the level scheme as a function of the 
magnetic field for $\lambda_D= 1.1\times 10^{-9} {\rm eV} {\rm cm}$. 
In this figure the active level for the 
same electron numbers of Fig.\ 2 are marked 
with thick dots and dashes. We note the clear correspondence of the
discontinuities in Fig.\ 2 with the crossings in Fig.\ 3, which 
correspond to changes in the precessional level.

In the rest of the work we present numerical results for the SO
precessional frequencies when the electron-electron Coulomb interaction
is added to the model. In this case we rely exclusively on the 
real-time simulation method since the perturbative treatment 
equivalent to Eq.\ (\ref{eq10}), known as the random-phase approximation,
becomes extremely demanding in the present context of spinors
without good angular momentum. 
Electronic exchange and correlation effects will be approximated within 
density-functional theory in the local-spin density approximation,
as in Refs.\ \cite{Pue99,Val02a}.
The Hamiltonian 
$h_{sp}({\bf r};\eta\eta')$ of Eq.\ (\ref{eq9}) is thus extended 
to include the dynamical terms
\begin{eqnarray}
\label{eq11}
V_H({\bf r};t) &=& \frac{e^2}{\kappa}
\int{d{\bf r}' {\rho({\bf r}';t)\over |{\bf r}'-{\bf r}|}}\; , \nonumber\\
V_{{\em xc}}({\bf r},\eta\eta';t) &=&
{\delta{E_{xc}}[\rho_{\eta\eta'}]\over
\delta\rho_{\eta\eta'}({\bf r};t)}\; ;
\end{eqnarray}
i.e., the Hartree and exchange-correlation contributions, respectively.
In Eq.\ (\ref{eq11}) we have used the spin density matrix
$\rho_{\eta\eta'}$, the total density 
$\rho$ and exchange-correlation energy 
$E_{xc}$, as well as the dielectric constant 
$\kappa=12.4$.

Figure 4 is the analog of Fig.\ 1 within LSDA. The time 
signal has a similar large period, but the lower period modulations
are manifestly different. Accordingly, the Fourier transform 
(upper panel) shows a similar low energy precessional mode but
the distribution of minor peaks is rather different. The dipole peaks 
of Fig.\ 1 are washed out and, instead, new excitations at 
$\hbar\omega\approx 1.7$ and $\approx$ 4 meV appear. 
As was discussed in Ref.\ \cite{Val02b} in the context of the 
far-infrared absorption, these excitations are
collective spin oscillations known as
dipole magnons.
Figure 2 also shows the LSDA numerical calculations for the higher
SO coupling constant. The characteristics of the precessional
mode discussed above are qualitatively retained within LSDA, 
although with the important difference that the discontinuity points 
are changed because of the interaction-induced orbital rearrangements. 

In general, along with the transverse magnetic field the system 
will be probed by an electric field. This modifies
the relative strength of the precessional mode with respect to the 
plasmon and magnon peaks of Figs.\ 1 and 4. 
We have checked this numerically by using an initial charge translation, 
simulating the effect of the electric field at $t=0$, simultaneously to 
the spinor rotation. 
The corresponding spectra display the same peaks of Fig.\ 1
for the non-interacting case and of Fig.\ 4 in LSDA, but with different 
heights. Therefore, only the strength, not the energy of the precessional 
mode depends on the coupling with the electric field.

In summary, the theory of electronic spin precession in GaAs 
quantum dots with SO coupling predicts a rich behaviour of the 
precessional frequencies with the electron number, the magnetic 
field and the intensity of the coupling. In this way, the spin precessional
channel reveals information not only about the SO coupling intensity
but also about the intrinsic level structure. 
It also opens the possibility to control the magnetic dynamical properties 
through the nanostructure parameters or, simply, by changing the
number of electrons in the quantum dot, which can actually be varied 
one by one in GaAs nanostructures using electric gates.
Although some of the precessional characteristics of GaAs dots   
we have discussed seem large enough to be experimentally  
accessible, the relevance for real samples of many effects beyond the 
ideal system considered here deserve more theoretical work. In particular,
we mention the possible dephasing mechanisms in dot samples, influence 
of the temperature on the electronic precession and role of the 
coupling with nuclear spins.  
 
This work was supported by Grant No.\ BFM2002-03241 from DGI (Spain),
and by COFINLAB from Murst (Italy).

\begin{figure}[f]
\centerline{\psfig{figure=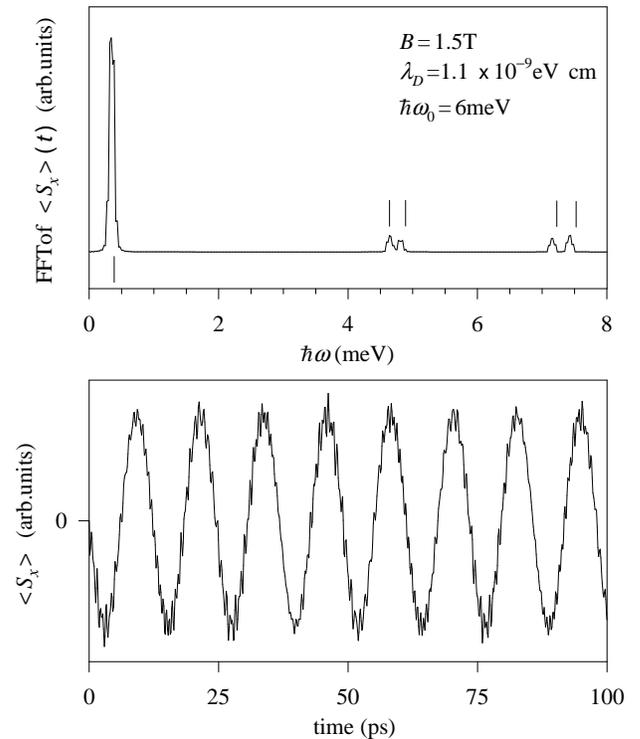,width=3.2in,clip=}}
\caption{Real time simulation of the spin evolution 
following an initial rotation with $x$ axis. Shown is the $x$-component
of total spin in time (lower) and energy (upper) domains.
The vertical bars in the upper panel indicate the analytical energies
with the Hamiltonian expanded to $O(\lambda_D^3)$.}
\end{figure}

\begin{figure}[f]
\centerline{\psfig{figure=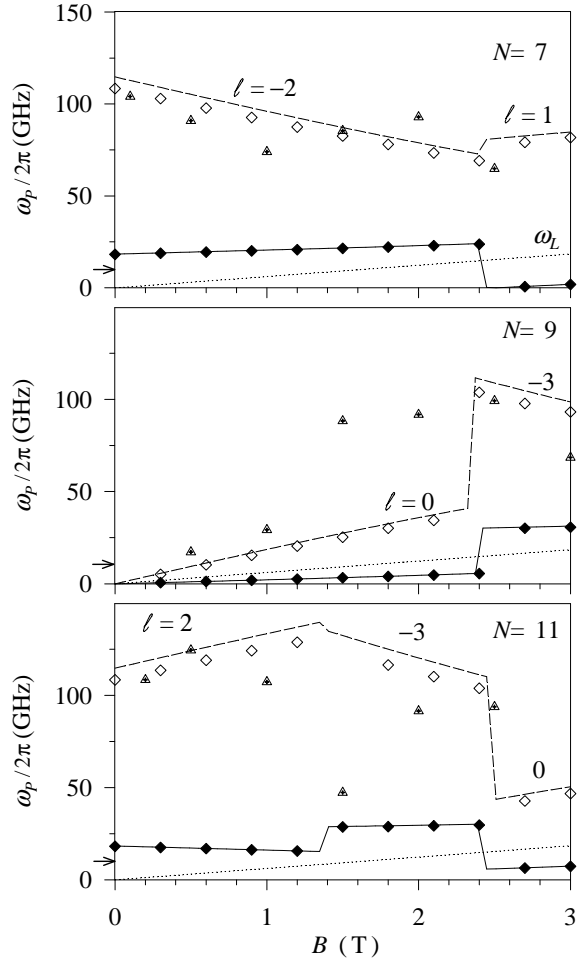,width=3.in,clip=}}
\caption{Systematics of SO precessional frequencies $\omega_{\cal P}$
as a function of magnetic field for two different values of the 
Dresselhaus parameter $\lambda_D$: in the analytical model (lines),
and from the numerical calculation without (diamonds)
and with Coulomb interaction (triangles with crosses).
Solid lines and symbols correspond to 
$\lambda_D=0.4\times10^{-9}\; {\rm eV}\;{\rm cm}$
while dashed lines, open symbols and crossed triangles to
$\lambda_D=1.1\times10^{-9}\; {\rm eV}\;{\rm cm}$.
The dotted line shows the Larmor frequency. Also indicated is the 
$\ell$ value of the precessionally active orbital in the analytical 
model, which at a given $B$ is the same for both $\lambda_D$ values.
The arrows on the vertical scale indicate the approximate lower
frequency than can be obtained from the time simulation window 
of 100 ps.}
\end{figure}

\begin{figure}[f]
\centerline{\psfig{figure=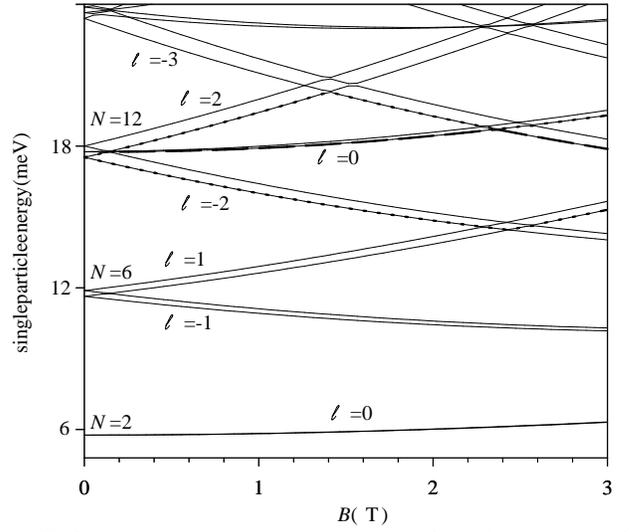,width=3.2in,clip=}}
\caption{Energy level scheme as a function of the magnetic field for the 
same dot of Figs.\ 1 and 2 with a SO parameter 
$\lambda_D= 1.1\times 10^{-9} {\rm eV} {\rm cm}$. 
The angular momentum for each level and the electron 
number at shell closures are indicated. The curves marked with thick
dots and dashes indicate the 
precessionally active level for the electron numbers $N=7$, 9 and 11.}
\end{figure}

\begin{figure}[f]
\centerline{\psfig{figure=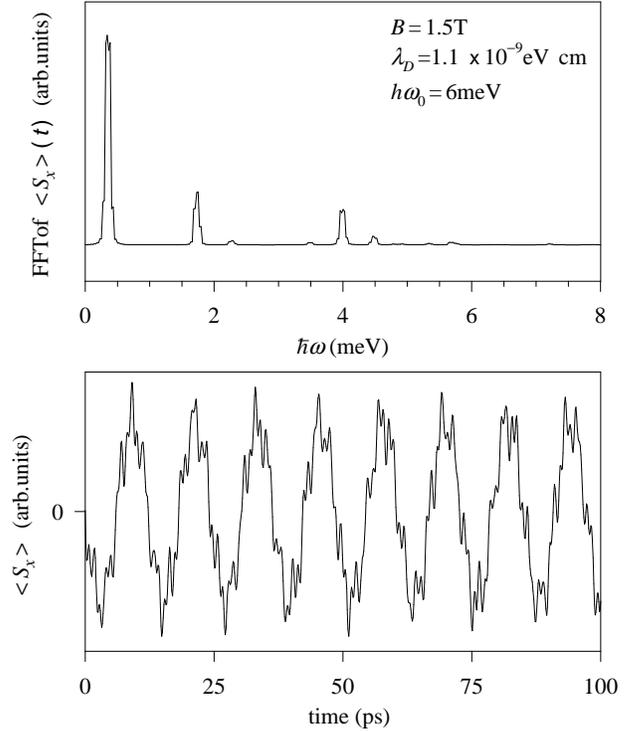,width=3.2in,clip=}}
\caption{Same as Fig.\ 1 for the case with Coulomb interaction between 
electrons.}
\end{figure}

\end{document}